\documentclass[12pt]{article}

\def\rmdj {d\llap{\raise 1.22ex\hbox
  {\vrule height 0.09ex width 0.315em}\kern 0.04em}}
\def\sldj {d\llap{\raise 1.22ex\hbox
  {\vrule height 0.09ex width 0.265em}}\rlap{\raise 1.22ex\hbox
  {\vrule height 0.09ex width 0.05em}}}
\def\itdj {d\llap{\raise 1.22ex\hbox
  {\vrule height 0.09ex width 0.2em}}\rlap{\raise 1.22ex\hbox
  {\vrule height 0.09ex width 0.06em}}}
\def\bfdj {d\llap{\raise 1.16ex\hbox
  {\vrule height 0.126ex width 0.308em}\kern 0.04em}}
\def\ttdj {\rlap{\kern 0.17em\raise 1.1ex\hbox
  {\vrule height 0.09ex width 0.295em}}d}
\def\scdj {\rlap{\kern 0.04em\raise 0.57ex\hbox
  {\vrule height 0.09ex width 0.20em}}d}
\def\sfdj {d\llap{\raise 1.22ex\hbox
  {\vrule height 0.10ex width 0.3em}\kern 0.02em}}
\def\dj{\ifcase\fam \rmdj \or \or \or \or \itdj \or \sldj
  \or \bfdj \or \ttdj \or \sfdj \or \scdj \else \rmdj \fi}
\def\rmDj {\rlap{\kern 0.05em\raise 0.76ex\hbox
  {\vrule height 0.10ex width 0.28em}}D}
\def\slDj {\rlap{\kern 0.1em\raise 0.76ex\hbox
  {\vrule height 0.1ex width 0.28em}}D}
\def\itDj {\rlap{\kern 0.145em\raise 0.76ex\hbox
  {\vrule height 0.1ex width 0.274em}}D}
\def\bfDj {\rlap{\kern 0.044em\raise 0.72ex\hbox
  {\vrule height 0.126ex width 0.287em}}D}
\def\ttDj {\rlap{\kern 0.02em\raise 0.67ex\hbox
  {\vrule height 0.105ex width 0.20em}}D}
\def\scDj {\rlap{\kern 0.08em\raise 0.73ex\hbox
  {\vrule height 0.12ex width 0.24em}}D}
\def\sfDj {\rlap{\kern 0.02em\raise 0.727ex\hbox
  {\vrule height 0.126ex width 0.26em}}D}
\def\Dj{\ifcase\fam \rmDj \or \or \or \or \itDj \or \slDj
  \or \bfDj \or \ttDj \or \sfDj \or \scDj \else \rmDj \fi}

\begin{document}

\date{}
\title{THE INVARIANT FORMULATION OF SPECIAL RELATIVITY, OR THE ''TRUE
TRANSFORMATIONS RELATIVITY,'' AND ELECTRODYNAMICS}
\author{Tomislav Ivezi\'{c} \\
Ru\ifcase\fam \rmdj \or \or \or \or \itdj \or \sldj
  \or \bfdj \or \ttdj \or \sfdj \or \scdj \else \rmdj \fi er Bo\v{s}%
kovi\'{c} Institute, Zagreb, Croatia \\
\textit{ivezic@rudjer.irb.hr\bigskip}}
\maketitle

\begin{abstract}
In the invariant approach to special relativity (SR), which we call the
''true transformations (TT)\ relativity,'' a physical quantity in the
four-dimensional spacetime is mathematically represented either by a true
tensor or equivalently by a coordinate-based geometric quantity comprising
both components and a basis. This invariant approach differs both from the
usual covariant approach, which mainly deals with the basis components of
tensors in a specific, i.e., Einstein's coordinatization of the chosen
inertial frame of reference,.and the usual noncovariant approach to SR in
which some quantities are not tensor quantities, but rather quantities from
''3+1'' space and time, e.g., the synchronously determined spatial length.
This noncovariant formulation of SR is called the ''apparent transformations
(AT)\ relativity.'' The principal difference between the ''TT relativity''
and the ''AT relativity'' arises from the difference in the concept of
sameness of a physical quantity for different observers. In the second part
of this paper we present the invariant formulation of electrodynamics with
the electromagnetic field tensor $F^{ab},$ and also the equivalent
formulation in terms of the four-vectors of the electric $E^{a}$ and
magnetic $B^{a}$ fields.\bigskip
\end{abstract}






\newpage \noindent \textbf{1.} \textbf{INTRODUCTION\medskip }

\noindent At present there are two main formulations of the classical
electrodynamics. The first one is the manifestly covariant formulation,
which deals with the component form, in Einstein's coordinatization, of
tensor quantities and tensor equations in the four-dimensional (4D)
spacetime, and where the electromagnetic field tensor $F^{\alpha \beta }$
(the component form; for the notation see the next section) contains all the
information about the electromagnetic field. (In the Einstein (''e'') \cite
{Einst} coordinatization Einstein's synchronization \cite{Einst} of distant
clocks and cartesian space coordinates $x^{i}$ are used in the chosen
inertial frame of reference (IFR).) The second one is the noncovariant
formulation dealing with the three-vectors (3-vectors), the electric field $%
\mathbf{E}$ and the magnetic field $\mathbf{B}$, and with equations
containing them. The whole latter formulation is given in ''3+1'' space and
time and was constructed by Maxwell before the appearance of Einstein's
theory of relativity \cite{Einst}. In \cite{ivezic,ive2} I have presented an
alternative covariant formulation of vacuum electrodynamics with the
electric and magnetic 4-vectors $E^{\alpha }$ and $B^{\alpha }$ (also the
component form), which is equivalent to the usual covariant formulation with
$F^{\alpha \beta }$. Recently \cite{tom1} the invariant formulation of
vacuum electrodynamics is presented with the electric and magnetic 4-vectors
$E^{a}$ and $B^{a}$ (true tensors), which is equivalent to the invariant
formulation with $F^{ab}$ (also true tensor). For the covariant formulation
of electrodynamics with $E^{\alpha }$ and $B^{\alpha }$ (the component form
of tensors in the ''e'' coordinatization ) see also \cite{Salas} and \cite
{Espos}. The covariant formulation with $F^{\alpha \beta }$ and the usual
formulation with the electric and magnetic 3-vectors $\mathbf{E}$ and $%
\mathbf{B}$ are generally considered to be equivalent. It is shown in \cite
{ivezic,ive2} (\cite{tom1}) that, contrary to the general opinion, there is
not the equivalence between covariant (invariant) formulations, either the
usual one with $F^{\alpha \beta }$ ($F^{ab}$), or equivalently the
alternative one with $E^{\alpha }$ and $B^{\alpha }$ ($E^{a}$ and $B^{a}$),
and the usual noncovariant formulation.

It seems that the work on the foundations of electromagnetic theory is again
in a continuos progress and I only quote two recent references, \cite{Dvoe}
and \cite{Chub}, which present an interesting part of this work.

In the first part of this paper a general discussion on the ''TT
relativity'' will be presented, and in the second part we consider the
invariant formulation of electrodynamics with $F^{ab},$ and $E^{a}$ and $%
B^{a}$. \bigskip

\noindent \textbf{2. A GENERAL DISCUSSION\ ON\ THE\ ''TT RELATIVITY\medskip }

\noindent In \cite{rohrl1} Rohrlich introduced the notions of the true
transformations (TT) and the apparent transformations (AT) of physical
quantities and emphasized the role of \emph{sameness} of a physical quantity
for different observers. This concept of sameness is also considered in the
same sense by Gamba \cite{gamba}.

The principal difference between the ''TT relativity,'' the usual covariant
formulation and the ''AT relativity'' stems from the difference in the
concept of \emph{sameness} of a physical system, i.e., of a physical
quantity, for different observers. This concept actually determines the
difference in what is to be understood as a relativistic theory.

In this paper we explore a formulation of special relativity (SR) that is
borrowed from general relativity. This is the formulation in which all
physical quantities (in the case when no basis has been introduced) are
described by \emph{true tensor fields}, that are defined on the 4D
spacetime, and that satisfy \emph{true tensor equations} representing
physical laws. The true tensors and true tensor equations are defined
without any reference frame. For the formulation of spacetime theories
without reference frames see, e.g., \cite{tamas}. When the coordinate system
is introduced the physical quantities are mathematically represented by the
coordinate-based geometric quantities (CBGQs) that satisfy the
coordinate-based geometric equations (CBGEs). \emph{The CBGQs contain both
the components and the basis one-forms and vectors }of the chosen IFR.
Speaking in mathematical language a tensor of type (k,l) is defined as a
linear function of k one-forms and l vectors (in old names, k covariant
vectors and l contravariant vectors) into the real numbers, see, e.g., \cite
{Wald,schutz,misner}. If a coordinate system is chosen in some IFR then, in
general, any tensor quantity can be reconstructed from its components and
from the basis vectors and basis 1-forms of that frame, i.e., it can be
written in a coordinate-based geometric language, see, e.g., \cite{misner}.
The symmetry transformations for the metric $g_{ab}$, i.e., the isometries,
leave the pseudo-Euclidean geometry of 4D spacetime of SR unchanged; if we
denote an isometry as $\Phi ^{*}$ then $(\Phi ^{*}g)_{ab}=g_{ab}.$ At the
same time they do not change the true tensor quantities, or equivalently the
CBGQs, in physical equations. Thus \emph{isometries} are what Rohrlich \cite
{rohrl1} calls \emph{the TT}. The formulation of SR that deals with true
tensor quantities and the TT is called the ''TT relativity.'' In the ''TT
relativity'' different coordinatizations of an IFR are allowed and they are
all equivalent in the description of physical phenomena. (An example of a
nonstandard synchronization, and thus nonstandard coordinatization as well,
that drastically differs from the Einstein synchronization is considered in
detail in \cite{tom1}.) In the ''TT relativity'' the concept of \emph{%
sameness} of a physical quantity is very clear. Namely \emph{the CBGQs
representing some 4D physical quantity in different relatively moving IFRs,
or in different coordinatizations of the chosen IFR, are all mathematically
equal. }Thus they are really \emph{the same quantity }for different
observers, or in different coordinatizations. We suppose that in the ''TT
relativity'' such 4D tensor quantities are well defined not only
mathematically but also \emph{experimentally}, as measurable quantities with
real physical meaning. \emph{The complete and well defined measurement from
the ''TT relativity'' viewpoint is such measurement in which all parts of
some 4D quantity are measured. }Different experiments that test SR are
discussed in \cite{tom2} and it is shown that all experiments, which are
complete from the ''TT relativity'' viewpoint, can be qualitatively and
quantitatively explained by the ''TT relativity,'' while some experiments
cannot be adequately explained by the ''AT relativity.''

In this paper I use the same convention with regard to indices as in \cite
{tom1,tom2}. Repeated indices imply summation. Latin indices $a,b,c,d,...$
are to be read according to\emph{\ the abstract index notation}, see \cite
{Wald}, Sec.2.4.; they ''...should be viewed as reminders of the number and
type of variables the tensor acts on, \emph{not} as basis components.'' They
designate geometric objects in 4D spacetime. Thus, e.g., $l_{AB}^{a}$ (a
distance 4-vector $l_{AB}^{a}=x_{B}^{a}-x_{A}^{a}$ between two events $A$
and $B$ with the position 4-vectors $x_{A}^{a}$ and $x_{B}^{a}$) and $%
x_{A,B}^{a}$ are (1,0) tensors and they are defined independently of any
coordinate system. Greek indices run from 0 to 3, while latin indices $%
i,j,k,l,...$ run from 1 to 3, and they both designate the components of some
geometric object in some coordinate system, e.g., $x^{\mu }(x^{0},x^{i})$
and $x^{\mu ^{\prime }}(x^{0^{\prime }},x^{i^{\prime }})$ are two coordinate
representations of the position 4-vector $x^{a}$ in two different inertial
coordinate systems $S$ and $S^{\prime }.$ The true tensor $x^{a}$ is then
represented as the CBGQs in different bases $\left\{ e_{\mu }\right\} $ in
an IFR $S$ and $\left\{ e_{\mu ^{\prime }}\right\} $ in a relatively moving
IFR $S^{\prime }$ as $x^{a}=x^{\mu }e_{\mu }=x^{\mu ^{\prime }}e_{\mu
^{\prime }},$ where, e.g., $e_{\mu }$ are the basis 4-vectors, $%
e_{0}=(1,0,0,0)$ and so on, and $x^{\mu }$ are the basis components when the
''e'' coordinatization is chosen in some IFR $S.$ Similarly the metric
tensor $g_{ab}$ denotes a tensor of type (0,2). The geometry of the
spacetime is generally defined by this metric tensor $g_{ab},$ which can be
expanded in a coordinate basis in terms of its components as $g_{ab}=g_{\mu
\nu }dx^{\mu }\otimes dx^{\nu },$ and where $dx^{\mu }\otimes dx^{\nu }$ is
an outer product of the basis 1-forms. Thus the geometric object $g_{ab}$ is
represented in the component form in an IFR $S,$ and in the ''e''
coordinatization, i.e., in the $\left\{ e_{\mu }\right\} $ basis, by the $%
4\times 4$ diagonal matrix of components of $g_{ab}$, $g_{\mu \nu
}=diag(-1,1,1,1),$ and this is usually called the Minkowski metric tensor.

It has to be noted that the ''TT relativity'' approach to SR differs not
only from the ''AT relativity'' approach but also from the usual covariant
approach. The difference lies in the fact that\emph{\ the usual covariant
approach} mainly deals with \emph{the basis components of tensors }%
(representing physical quantities) and the equations of physics are written
out\emph{\ in the component form.} Mathematically speaking the concept of a
tensor in the usual covariant approach is defined entirely in terms of\emph{%
\ the transformation properties of its components relative to some
coordinate system. }Obviously in the usual covariant approach (including
\cite{rohrl1} and \cite{gamba})\emph{\ the basis components of a true
tensor, }or equivalently of a CBGQ, that are determined in different IFRs
(or in different coordinatizations), are considered to be \emph{the same
quantity }for different observers. Although the basis components of a true
tensor refer to the same tensor quantity they, in fact, are not the same
quantity. They depend on the chosen reference frame and the chosen
coordinatization of that reference frame. Thus the basis components are the
coordinate quantities.

In contrast to the TT \emph{the AT }are not the transformations of spacetime
tensors and they do not refer to the same 4D quantity. Thus they are not
isometries and they \emph{refer exclusively to the component form of tensor
quantities and in that form they transform only some components of the whole
tensor quantity. }In fact, depending on the used AT, only a part of a 4D
tensor quantity is transformed by the AT. Such a part of a 4D quantity, when
considered in different IFRs (or in different coordinatizations of some IFR)
corresponds to different quantities in 4D spacetime. An example of\ the AT
is the AT of the synchronously defined spatial length \cite{Einst}, i.e.,
the Lorentz ''contraction.'' It is shown in \cite{rohrl1,gamba}, and more
exactly in \cite{tom1}, that the Lorentz ''contraction'' is an AT. The
spatial or temporal distances taken alone are not well defined quantities in
4D spacetime. Further it is shown in \cite{ivezic,ive2}, and more exactly in
\cite{tom1}, that the conventional transformations of the electric and
magnetic 3-vectors $\mathbf{E}$ and $\mathbf{B}$ (see, e.g., \cite{jacks}
Sec.11.10) are also the AT. The formulation of SR which uses the AT we call
the ''AT relativity.'' An example of such formulation is Einstein's
formulation of SR which is based on his two postulates and which deals with
different AT. Thus in the ''AT relativity'' quantities connected by an AT,
e.g., two spatial lengths connected by the Lorentz contraction, are
considered to be the same quantity for different observers. However, as
explicitly shown in \cite{tom1}, the quantities connected by an AT are not
well defined quantities in 4D spacetime and, actually, they correspond to
different quantities in 4D spacetime.

(In the following we shall also need the expression for the covariant 4D
Lorentz transformations $L^{a}{}_{b}$, \emph{which is independent of the
chosen coordinatization of reference frames }(see \cite{Fahn}, \cite{ive2}
and \cite{tom1}). It is
\begin{equation}
L^{a}{}_{b}\equiv L^{a}{}_{b}(v)=g^{a}{}_{b}-\frac{2u^{a}v_{b}}{c^{2}}+\frac{%
(u^{a}+v^{a})(u_{b}+v_{b})}{c^{2}(1+\gamma )},  \label{fah}
\end{equation}
where $u^{a}$ is the proper velocity 4-vector of a frame $S$ with respect to
itself, $u^{a}=cn^{a},$ $n^{a}$ is the unit 4-vector along the $x^{0}$ axis
of the frame $S,$ and $v^{a}$ is the proper velocity 4-vector of $S^{\prime
} $ relative to $S.$ Further $u\cdot v=u^{a}v_{a}$ and $\gamma =-u\cdot
v/c^{2}.$ In the Einstein coordinatization $L^{a}{}_{b}$ is represented by $%
L^{\mu }{}_{\nu },$ the usual expression for pure Lorentz transformation
which connects two coordinate representations, basis components $x^{\mu },$ $%
x^{\mu ^{\prime }}$ of a given event. $x^{\mu },$ $x^{\mu ^{\prime }}$ refer
to two relatively moving IFRs (with the Minkowski metric tensors) $S$ and $%
S^{\prime },$
\begin{eqnarray}
x^{\mu ^{\prime }} &=&L^{\mu ^{\prime }}{}_{\nu }x^{\nu },\qquad
\,L^{0^{\prime }}{}_{0}=\gamma ,\quad L^{0^{\prime }}{}_{i}=L^{i^{\prime
}}{}_{0}=-\gamma v^{i}/c,  \nonumber \\
L^{i^{\prime }}{}_{j} &=&\delta _{j}^{i}+(\gamma -1)v^{i}v_{j}/v^{2},
\label{lorus}
\end{eqnarray}
where $v^{\mu }\equiv dx^{\mu }/d\tau =(\gamma c,\gamma v^{i}),$ $d\tau
\equiv dt/\gamma $ and $\gamma \equiv (1-v^{2}/c^{2})^{1/2}$. Since $g_{\mu
\nu }$ is a diagonal matrix the space $x^{i}$ and time $t$ $(x^{0}\equiv ct)$
parts of $x^{\mu }$ do have their usual meaning.)

As already mentioned different experiments that test SR are discussed in
\cite{tom2}. In numerous papers and textbooks it is considered that the
experiments on the length contraction and the time dilatation test SR, but
the discussion from \cite{tom2} shows that such an interpretation of the
experiments refers exclusively to - the ''AT relativity,'' and not to - the
''TT relativity.'' When SR is understood as the theory of 4D spacetime with
pseudo-Euclidean geometry then instead of the Lorentz contraction and the
dilatation of time one has to consider the 4D tensor quantities, the
spacetime length $l$, $l=(g_{ab}l^{a}l^{b})^{1/2}$, or the distance 4-vector
$l_{AB}^{a}=x_{B}^{a}-x_{A}^{a}.$ Namely in the ''TT relativity'' the
measurements in different IFRs (and different coordinatizations) have to
refer to the same 4D tensor quantity, i.e., to a CBGQ. In the chosen IFR and
the chosen coordinatization the measurement of some 4D quantity has to
contain the measurements of all parts (all the basis components) of such a
quantity. However in almost all experiments that refer to SR only the
quantities belonging to the ''AT relativity'' were measured. From the ''TT
relativity'' viewpoint such measurements are incomplete, since only some
parts of a 4D quantity, not all, are measured. It is shown in \cite{tom2}
that the ''TT relativity'' theoretical results agree with all experiments
that are complete from the ''TT relativity'' viewpoint, i.e., in which all
parts of the considered tensor quantity are measured in the experiment.
However the ''AT relativity'' results agree only with some of the examined
experiments (and this agreement exists only for the specific
coordinatization, i.e., the ''e'' coordinatization. Moreover the agreement
of the ''AT relativity'' and the experiments is, in fact, an ''apparent''
agreement, which is usually obtained by means of an incorrect treatment of
4D quantities. This is explicitly shown in \cite{tom2} for some of the
well-known experiments: the ''muon'' experiment, the Michelson-Morley type
experiments, the Kennedy-Thorndike type experiments and the Ives-Stilwell
type experiments.

In this paper we only give a short discussion of the Michelson-Morley
experiment and for the details see \cite{tom2}. In the Michelson-Morley
experiment two light beams emitted by one source are sent, by half-silvered
mirror $O$, in orthogonal directions. These partial beams of light traverse
the two equal (of the length $L$) and perpendicular arms $OM_{1}$
(perpendicular to the motion) and $OM_{2}$ (in the line of motion) of
Michelson's interferometer and the behavior of the interference fringes
produced on bringing together these two beams after reflection on the
mirrors $M_{1}$ and $M_{2}$ is examined. The Earth frame is the rest frame
of the interferometer, i.e.,\textbf{\ }it is the $S$\ frame, while the $%
S^{\prime }$\ frame is the (preferred) frame\ in which the interferometer is
moving at velocity $\mathbf{v.}$

In the Michelson-Morley experiment the traditional, ''AT relativity,''
derivation of the fringe shift $\bigtriangleup N$ deals only with the
calculation, in the ''e'' coordinatization, of $t_{1}$ and $t_{2}$ (in $S$
and $S^{\prime }$)$,$ which are the times required for the complete trips $%
OM_{1}O$ and $OM_{2}O$ along the arms of the Michelson-Morley
interferometer; . The null fringe shift obtained with such calculation is
only in an ''apparent,'' not true, agreement with the observed null fringe
shift, since this agreement was obtained by an incorrect procedure. Namely
it is supposed in such derivation that, e.g., $t_{1}$ and $t_{1}^{\prime }$
refer to the same quantity measured by the observers in relatively moving
IFRs $S$ and $S^{\prime }$ that are connected by the Lorentz transformation.
However, as shown in \cite{tom1,tom2}, the relation for the time dilatation $%
t_{1}^{\prime }=\gamma t_{1},$ which is used in the usual explanation of the
Michelson-Morley experiment, is not the Lorentz transformation of some 4D
quantity, and, see \cite{tom2}, $t_{1}^{\prime }$ ($t_{1}^{\prime
}=2L/c(1-v^{2}/c^{2})^{1/2}$) and $t_{1}$ ($t_{1}=2L/c$) do not correspond
to the same 4D quantity considered in $S^{\prime }$ and $S$ respectively but
to different 4D quantities.

Our ''TT relativity'' calculation, in contrast to the ''AT relativity''
calculation, deals always with the true tensor quantities or the CBGQs; in
the Michelson-Morley experiment it is the phase of a light wave
\begin{equation}
\phi =k^{a}g_{ab}l^{b},  \label{phase}
\end{equation}
where $k^{a}$ is the propagation 4-vector, $g_{ab}$ is the metric tensor and
$l^{b}$ is the distance 4-vector. All quantities in (\ref{phase}) are true
tensor quantities and thus (\ref{phase}) is written without any reference
frame. These quantities can be written in the coordinate-based geometric
language and, e.g., the decompositions of $k^{a}$ in $S$ and $S^{\prime }$
and in the ''e'' coordinatization are
\begin{equation}
k^{a}=k^{\mu }e_{\mu }=k^{\mu ^{\prime }}e_{\mu ^{\prime }},  \label{kad}
\end{equation}
where the basis components $k^{\mu }$ of the CBGQ are transformed by $L^{\mu
^{\prime }}{}_{\nu }$ (\ref{lorus}), while the basis vectors $e_{\mu }$ are
transformed by the inverse transformation $(L^{\mu ^{\prime }}{}_{\nu
})^{-1}=L^{\mu }{}_{\nu ^{\prime }}.$ By the same reasoning the phase $\phi $
(\ref{phase}) is given in the coordinate-based geometric language as
\begin{equation}
\phi =k^{\mu }g_{\mu \nu }\,l^{\nu }=k^{\mu ^{\prime }}g_{\mu \nu }\,l^{\nu
^{\prime }}.  \label{pha2}
\end{equation}
As shown in \cite{tom2} the ''TT relativity'' calculations yields the
observed null fringe shift and that result holds for all IFRs and all
coordinatizations.

In addition, it is shown in \cite{tom2} that the usual ''AT relativity''
actually deals only with the part $k^{0}l_{0}$ (i.e., $\omega t$) of the
whole phase $\phi ,$ (\ref{phase}) or (\ref{pha2}). This contribution $%
k^{0}l_{0}$ is considered in the interferometer rest frame $S,$ while in the
$S^{\prime }$ frame, in which the interferometer is moving, the usual ''AT
relativity'' takes into account only the contribution $k^{0}l_{0^{\prime }}$
(i.e., $\omega t^{\prime }$); the $k^{0}$ (i.e., $\omega $) factor is taken
to be the same in $S$ and $S^{\prime }$ frames. Thus in the usual ''AT
relativity'' two different quantities $k^{0}l_{0}$ and $k^{0}l_{0^{\prime }}$
(only the parts of the phase (\ref{phase}) or (\ref{pha2})) are considered
to be the same 4D quantity for observers in $S$ and $S^{\prime }$ frames,
and these quantities are considered to be connected by the Lorentz
transformation. Such an incorrect procedure then caused an apparent (not
true) agreement of the traditional analysis with the results of the
Michelson-Morley experiment. Since only a part of the whole phase $\phi $ (%
\ref{phase}) or (\ref{pha2}) is considered the traditional result is
synchronization, i.e., coordinatization, dependent result.

Driscoll \cite{drisc} \emph{improved} the traditional ''AT relativity''
derivation of the fringe shift taking into account the changes in
frequencies due to the Doppler effect. (Recall that in the traditional
approach $\omega $ is the same in $S$\ and $S^{\prime }$.) The improved ''AT
relativity'' calculation of the fringe shift from \cite{drisc} finds a
''surprising'' non-null fringe shift. It is shown in \cite{tom2} that the
non-null theoretical result for the fringe shift from \cite{drisc} is easily
obtained from our ''TT relativity'' approach taking only the product $%
k^{0^{\prime }}l_{0^{\prime }}$\ in the calculation of the increment of
phase $\phi ^{\prime }$\ in $S^{\prime }$ in which the apparatus is moving.
Thus again as in the usual ''AT relativity'' calculation two different
quantities $k^{0}l_{0}$ and $k^{0^{\prime }}l_{0^{\prime }}$ (only the parts
of the phase (\ref{phase}) or (\ref{pha2})) are considered to be the same 4D
quantity for observers in $S$ and $S^{\prime }$ frames, and consequently
that these two quantities are connected by the Lorentz transformation. Since
only a part $k^{0^{\prime }}l_{0^{\prime }}$ of the whole 4D tensor quantity
$\phi $ (\ref{phase}) or (\ref{pha2}) is considered the non-null fringe
shift can be shown to be quite different in another coordinatization, see
\cite{tom2}.

The same conclusions can be drawn for the Kennedy-Thorndike type
experiments, and for the modern laser versions of both, the Michelson-Morley
and the Kennedy-Thorndike type experiments, see \cite{tom2}.

This short consideration illustrates the main differences in the
interpretation of the well-known experiments from the point of view of the
traditional ''AT relativity'' and from the viewpoint of the ''TT
relativity.'' \bigskip

\noindent \textbf{3. THE INVARIANT FORMULATION\ OF\ ELECTRODYNAMICS\ WITH}\ $%
F^{ab}\medskip $

\noindent Let us now apply the above general consideration of the invariant
formulation of SR to the electrodynamics.

The usual covariant Maxwell equations with $F^{\alpha \beta }$ and its dual $%
^{*}F^{\alpha \beta }$
\begin{equation}
\partial _{\alpha }F^{a\beta }=-j^{\beta }/\varepsilon _{0}c,\quad \partial
_{\alpha }\ ^{*}F^{\alpha \beta }=0,  \label{maxco}
\end{equation}
where $^{*}F^{\alpha \beta }=-(1/2)\varepsilon ^{\alpha \beta \gamma \delta
}F_{\gamma \delta }$ and $\varepsilon ^{\alpha \beta \gamma \delta }$ is the
totally skew-symmetric Levi-Civita pseudotensor, are actually the equations
in the ''e'' coordinatization for basis components in a chosen IFR. We first
show how these equations for the basis components are derived from the true
tensor equations (when no basis has been introduced). The true tensor
equations can be written in the abstract index notation as
\begin{equation}
\nabla ^{a}F_{ab}=-j_{b}/\varepsilon _{0}c,\quad \varepsilon ^{abcd}\nabla
_{b}F_{cd}=0,  \label{mxt3}
\end{equation}
where $\nabla _{b}$ is the derivative operator (sometimes called the
covariant derivative), see, e.g., \cite{Wald}. The tensor equation (\ref
{mxt3}) can be written in the following form
\begin{equation}
(-g)^{-1/2}\partial _{a}((-g)^{1/2}F^{ab})=-j^{b}/\varepsilon _{0}c,\quad
\varepsilon ^{abcd}\partial _{b}F_{cd}=0,  \label{maxten}
\end{equation}
where $g$ is the determinant of the metric tensor $g_{ab}$ and $\partial
_{a} $ is an ordinary derivative operator. When some coordinatization is
chosen in a specific IFR $S$, e.g., the ''e'' coordinatization, then the
relations (\ref{maxten}) can be written in the coordinate-based geometric
language as the equations that contain the basis vectors as well,
\begin{equation}
\partial _{\alpha }F^{a\beta }e_{\beta }=-(1/\varepsilon _{0}c)j^{\beta
}e_{\beta },\quad \partial _{\alpha }\ ^{*}F^{\alpha \beta }e_{\beta }=0.
\label{maco1}
\end{equation}
(We remark that (\ref{maco1}) follows from (\ref{maxten}) for those
coordinatizations for which the basis vectors are constant, e.g., the ''e''
coordinatization.) From (\ref{maco1}), which contain the basis (1,0) tensors
(4-vectors), one finds the already written equations for basis components (%
\ref{maxco}); every equation in (\ref{maco1}) is the equality of two tensors
of the same type, two 4-vectors, and if two 4-vectors are equal then the
corresponding components are equal, and that holds in all bases. In many
treatments only the covariant Maxwell equations (\ref{maxco}) for the basis
components are used forgetting that they are obtained from the tensor
equations (\ref{maxten}) or (\ref{maco1}).

Similarly one finds from (\ref{maxten}) the equations corresponding to (\ref
{maco1}) and to (\ref{maxco}) but in the $\left\{ e_{\mu ^{\prime }}\right\}
$ basis, i.e., in the $S^{\prime }$ frame and in the ''e'' coordinatization,
by replacing the unprimed quantities with the primed ones.

>From this consideration some important conclusions can be derived regarding
the mathematical form of the physical laws in the ''TT relativity.'' From
the mathematical viewpoint the (1,0) tensor quantity $(-g)^{-1/2}\partial
_{a}((-g)^{1/2}F^{ab})$ can be written in the coordinate-based geometric
language in the ''e'' cordinatization, and in $S$ as $\partial _{\alpha
}F^{a\beta }e_{\beta },$ while in $S^{\prime }$ as $\partial _{\alpha
^{\prime }}F^{a^{\prime }\beta ^{\prime }}e_{\beta ^{\prime }},$ where all
primed quantities (including the basis vectors) are obtained by the TT,
i.e., by the Lorentz transformation $L^{\mu }{}_{\nu ,e}$ (\ref{lorus}) from
the corresponding unprimed quantities. Thus
\begin{equation}
(-g)^{-1/2}\partial _{a}((-g)^{1/2}F^{ab})=\partial _{\alpha }F^{a\beta
}e_{\beta }=\partial _{\alpha ^{\prime }}F^{a^{\prime }\beta ^{\prime
}}e_{\beta ^{\prime }},  \label{mxj}
\end{equation}
which shows that \emph{the equalities in (\ref{mxj}) refer to the same
quantity in 4-D spacetime.} Analogously, the mathematics yields for the
(1,0) tensor (4-vector) $-j^{b}/\varepsilon _{0}c$ the relations
\begin{equation}
-j^{b}/\varepsilon _{0}c=-(1/\varepsilon _{0}c)j^{\beta }e_{\beta
}=-(1/\varepsilon _{0}c)j^{\beta ^{\prime }}e_{\beta ^{\prime }}.
\label{mx2}
\end{equation}
A similar analysis can be applied to the second Maxwell equation in (\ref
{maxten}).

\emph{The physical laws} expressed as tensor equations, e.g., (\ref{maxten}%
), or equivalently as CBGEs, for example, (\ref{maco1}), \emph{set up the
connection between two geometric quantities}, in this case, two 4-vectors,
that are given by equations (\ref{mxj}) and (\ref{mx2}). \emph{The
experiments in which all parts of tensor quantities are measured then play
the fundamental role in deciding about the validity of some physical law
mathematically expressed as tensor equation. }We see from the equations (\ref
{mxj}) and (\ref{mx2}) that \emph{when the physical laws are expressed as
tensor, geometric, equations} (\ref{maxten}) or (\ref{maco1}) \emph{then
these equations are invariant upon the Lorentz transformations. }It is not
so for the equations in the component form, e.g., (\ref{maxco}). Of course
the covariance of physical equations, when they are written in the component
form, is a simple consequence of the invariance of tensor quantities, or
equivalently, of the CBGQs, upon the mentioned TT, that is upon the
isometries. \emph{The invariance of physical laws, that are expressed as
tensor equations, or equivalently as the CBGEs, means that all physical
phenomena proceed in the same way (taking into account the corresponding
initial and boundary conditions) in different IFRs. Thus there is no
physical difference between these frames, what automatically embodies the
principle of relativity. We remark that in the ''TT relativity'' there is no
need to postulate the principle of relativity as a fundamental law. It is
replaced by the requirement that the physical laws must be expressed as
tensor equations (or equivalently as the CBGEs) in the 4D spacetime.}

This consideration is used in \cite{tom1} to derive an important result,
i.e., to show that, contrary to the general belief,\emph{\ the usual Maxwell
equations with }$E_{i}$\emph{\ and }$B_{i},$ \emph{or with the 3-vectors }$%
\mathbf{E}$\emph{\ and }$\mathbf{B,}$\ \emph{are not equivalent to the
tensor equations (\ref{maxten}), i.e., to the CBGEs (\ref{maco1}).} Further
it is explicitly shown in \cite{tom1} that the conventional transformations
for $\mathbf{E}$ and $\mathbf{B}$ (see, e.g., \cite{jacks} Sec.11.10)
actually connect different quantities in 4D spacetime, and thus that they
are not the TT but the AT. \bigskip

\noindent \textbf{4. THE INVARIANT FORMULATION\ OF\ ELECTRODYNAMICS\ WITH}\ $%
E^{a}$ \textbf{AND} $B^{a}\medskip $

\noindent In this section we present the formulation of electrodynamics
introducing the 4-vectors $E^{a}$ and $B^{a}$ instead of the usual 3-vectors
$\mathbf{E}$ and $\mathbf{B.}$ The Maxwell equations are formulated as
tensor equations with $E^{a}$ and $B^{a},$ which are equivalent to the
tensor Maxwell equations with $F^{ab},$ (\ref{mxt3}) or (\ref{maxten})$.$ We
define the electric and magnetic fields by the relations
\begin{equation}
E_{a}=(1/c)F_{ab}v^{b},\quad B^{a}=-(1/2c^{2})\varepsilon ^{abcd}v_{b}F_{cd}.
\label{veef}
\end{equation}
The $E^{a}$ and $B^{a}$ are the electric and magnetic field 4-vectors
measured by an observer moving with 4-velocity $v^{a}$ in an arbitrary
reference frame, $v^{a}v_{a}=-c^{2},$ and $\varepsilon ^{abcd}$ is the
totally skew-symmetric Levi-Civita pseudotensor (density). These fields
satisfy the conditions $v_{a}E^{a}=v_{b}B^{b}=0,$ which follow from the
definitions (\ref{veef}) and the antisymmetry of $F_{ab}.$ In the usual
treatments (see, e.g., \cite{Wald}, \cite{sonego}, \cite{vanzel}) the
tensors $E^{a}$ and $B^{a}$ are introduced in the curved spacetimes or
noninertial frames, but at the same time the usual Maxwell equations with
the 3-vectors $\mathbf{E}$ and $\mathbf{B}$ are considered to be valid in
the IFRs. One gets the impression that $E^{a}$ and $B^{a}$ are considered
only as useful mathematical objects, while the real physical meaning is
associated with the 3-vectors $\mathbf{E}$ and $\mathbf{B.}$ Our results
obtained in \cite{tom1} and in Sec. 2. imply that it is necessary to use the
4-vectors $E^{a}$ and $B^{a}$ in IFRs as well. This means that the tensor
quantities $E^{a}$ and $B^{a}$ do have the real physical meaning and not the
3-vectors $\mathbf{E}$ and $\mathbf{B.}$ The inverse relation connecting the
$E^{a},B^{a}$ and the tensor $F_{ab}$ is
\begin{equation}
F_{ab}=(1/c)(v_{a}E_{b}-v_{b}E_{a})+\varepsilon _{abcd}v^{c}B^{d}.\
\label{vezFE}
\end{equation}
The tensor Maxwell equations with $E^{a},B^{a}$ in the curved spacetimes are
derived, e.g., in \cite{sonego}. Here we specify them to the IFRs, but in
such a way that they remain valid for different coordinatizations of the
chosen IFR. First we write the tensor Maxwell equations (\ref{maxten}) with $%
F^{ab}$ as the CBGEs (\ref{maco1}). Then we also write the equation (\ref
{vezFE}) in the coordinate-based geometric language and the obtained
equation substitute into (\ref{maco1}). This procedure yields
\begin{eqnarray}
\partial _{\alpha }(\delta _{\quad \mu \nu }^{\alpha \beta }v^{\mu }E^{\nu
}+c\varepsilon ^{\alpha \beta \mu \nu }B_{\mu }v_{\nu })e_{\beta }
&=&-(j^{\beta }/\varepsilon _{0})e_{\beta },  \nonumber \\
\partial _{\alpha }(\delta _{\quad \mu \nu }^{\alpha \beta }v^{\mu }B^{\nu
}+(1/c)\varepsilon ^{\alpha \beta \mu \nu }v_{\mu }E_{\nu })e_{\beta } &=&0,
\label{maeb}
\end{eqnarray}
where $E^{\alpha }$ and $B^{\alpha }$ are the basis components of the
electric and magnetic field 4-vectors $E^{a}$ and $B^{a}$ measured by a
family of observers moving with 4-velocity $v^{\alpha }$, and $\delta
_{\quad \mu \nu }^{\alpha \beta }=\delta _{\,\,\mu }^{\alpha }\delta
_{\,\,\nu }^{\beta }-\delta _{\,\,\nu }^{\alpha }\delta _{\,\,\mu }^{\beta
}. $ The equations (\ref{maeb}) correspond in the $E^{a},$ $B^{a}$ picture
to the equations (\ref{maco1}) in the $F^{ab}$ picture. From the relations (%
\ref{maeb}) we again find the covariant Maxwell equations for the basis
components (without the basis vectors $e_{\beta }$), which were already
presented in \cite{ivezic}, \cite{ive2} and \cite{tom1}.
\begin{eqnarray}
\partial _{\alpha }(\delta _{\quad \mu \nu }^{\alpha \beta }v^{\mu }E^{\nu
}+c\varepsilon ^{\alpha \beta \mu \nu }B_{\mu }v_{\nu }) &=&-(j^{\beta
}/\varepsilon _{0}),  \nonumber \\
\partial _{\alpha }(\delta _{\quad \mu \nu }^{\alpha \beta }v^{\mu }B^{\nu
}+(1/c)\varepsilon ^{\alpha \beta \mu \nu }v_{\mu }E_{\nu }) &=&0.
\label{ma4}
\end{eqnarray}
(It has to be mentioned that the component form of Maxwell equations, (\ref
{ma4}), was also presented in \cite{Salas}, and with $j^{\beta }=0$ in \cite
{Espos}$.$ However in \cite{Espos} the physical meaning of $v^{\alpha }$ is
unspecified - it is any unitary 4-vector. The reason for such choice of $%
v^{\alpha }$ in \cite{Espos} is that there $E^{\alpha }$ and $B^{\alpha }$
are introduced as the ''auxiliary fields,'' while $\mathbf{E}$ and $\mathbf{B%
}$ are considered as the physical fields. In our ''invariant'' approach with
$E^{a}$ and $B^{a}$ the situation is just the opposite; $E^{a}$ and $B^{a}$
are the real physical fields, which are correctly defined and measured in 4D
spacetime, while the 3-vectors $\mathbf{E}$ and $\mathbf{B}$ are not
correctly defined in 4D spacetime from the ''TT viewpoint.'' The equations (%
\ref{ma4}) for basis components correspond to the covariant Maxwell
equations for basis components (\ref{maxco}). Instead of to work with $%
F^{ab} $- formulation, (\ref{maco1}) and (\ref{maxco}), one can equivalently
use the $E^{a},B^{a}$ formulation with (\ref{maeb}) and (\ref{ma4}). For the
given sources $j^{a}$ one could solve these equations and find the general
solutions for $E^{a}$ and $B^{a}.$ \bigskip

\noindent \textbf{4.1 The comparison of Maxwell's equations with} $\mathbf{E}
$ \textbf{and} $\mathbf{B}$ \textbf{and those with} $E^{a}$ \textbf{and} $%
B^{a}\medskip $

\noindent The comparison of this invariant approach with $E^{a}$ and $B^{a}$
and the usual noncovariant approach with the 3-vectors $\mathbf{E}$ and $%
\mathbf{B}$ is possible in the ''e'' coordinatization. If one considers the
''e'' coordinatization and takes that in an IFR $S$ the observers who
measure the basis components $E^{\alpha }$ and $B^{\alpha }$ are at rest,
i.e., $v^{\alpha }=(c,\mathbf{0})$, then $E^{0}=B^{0}=0$, and one can derive
from the covariant Maxwell equations (\ref{ma4}) for the basis components $%
E^{\alpha }$ and $B^{\alpha }$ the Maxwell equations which contain only the
space parts $E^{i}$ and $B^{i}$ of $E^{\alpha }$ and $B^{\alpha }$, e.g.,
from the first covariant Maxwell equation in (\ref{ma4}) one easily finds $%
\partial _{i}E^{i}=j^{0}/\varepsilon _{0}c$. We see that the Maxwell
equations obtained in such a way from the Maxwell equations (\ref{maeb}), or
(\ref{ma4}), are of the same form as the usual Maxwell equations with $%
\mathbf{E}$ and $\mathbf{B}$. From the above consideration one concludes
that all the results obtained in a given IFR $S$ from the usual Maxwell
equations with $\mathbf{E}$ and $\mathbf{B}$ remain valid in the formulation
with the 4-vectors $E^{a}$ and $B^{a}$ \emph{(in the ''e''
coordinatization), but only for the observers who measure the fields }$E^{a}$%
\emph{\ and }$B^{a}$\emph{\ and are at rest in the considered IFR.} Then for
such observers the components of $\mathbf{E}$ and $\mathbf{B}$, which are
not well defined quantities in the ''TT relativity,'' can be simply replaced
by the space components of the 4-vectors $E^{a}$ and $B^{a}$ (in the ''e''
coordinatization). It has to be noted that just such observers were usually
considered in the conventional formulation with the 3-vectors $\mathbf{E}$
and $\mathbf{B.}$ However, the observers who are at rest in some IFR $S$
cannot remain at rest in another IFR $S^{\prime }$ moving with $V^{\alpha }$
relative to $S$. Hence in $S^{\prime }$ this simple replacement does not
hold; in $S^{\prime }$ one cannot obtain the usual Maxwell equations with
the 3-vectors $\mathbf{E}^{\prime }$ and $\mathbf{B}^{\prime }$ from the
transformed covariant Maxwell equations with $E^{\alpha ^{\prime }}$ and $%
B^{\alpha ^{\prime }}$.

Some important experimental consequences of the ''TT relativity'' approach
to electrodynamics have been derived in \cite{ive2}. They are the existence
of the spatial components $E^{i}$ of $E^{a}$ outside a current-carrying
conductor for the observers (who measure $E^{a}$) at rest in the rest frame
of the wire, and the existence of opposite (invariant) charges on opposite
sides of a square loop with current, both when \emph{the loop is at rest}
and when it is moving.

The similar external second-order electric fields from steady currents in a
conductor at rest are also predicted in, e.g., \cite{assis}. But this
prediction is made on the basis of Weber's theory and thus the theory from
\cite{assis} is an action-at-a-distance theory. \bigskip

\noindent \textbf{5. SUMMARY\ AND\ CONCLUSIONS}\medskip

\noindent In this paper we have presented the invariant (true tensor)
formulation of SR. This ''TT relativity'' is compared with the usual
covariant approach to SR and with the usual ''AT relativity'' formulation,
i.e., with the original Einstein's formulation.

The principal concept that makes distinction between the ''TT relativity''
formulation, the usual covariant formulation and the ''AT relativity''
formulation of SR is the concept of \emph{sameness} of a physical quantity
for different observers. In the ''TT relativity'' the same quantity for
different observers is the true tensor quantity, or equivalently the CBGQ,
only one quantity in 4D spacetime.

In the usual covariant approach one deals with the basis components of
tensors and with the equations of physics written out in the component form,
and all is mainly done in the ''e'' coordinatization. There one considers
that the basis components, e.g., $l^{\mu }$ and $l^{\mu ^{\prime }}$,
represent the same quantity for different observers. These quantities, in
fact, are not equal $l^{\mu }\neq l^{\mu ^{\prime }},$ but they only refer
to the same tensor quantity $l_{AB}^{a}$. If only one coordinatization is
\emph{always} used, usually the ''e'' coordinatization, then the
conventional covariant approach can be applied. However the physics must not
depend on the chosen coordinatization, which means that the theory has to be
formulated in the manner that does not depend on the choice of some specific
coordinatization. The Einstein coordinatization is nothing more physical but
any other permissible coordinatization. This requirement is fulfilled in the
''TT relativity.''

In the ''AT relativity'' one does not deal with tensor quantities but with
quantities from ''3+1'' space and time, e.g., the synchronously determined
spatial lengths, or the temporal distances taken alone. The AT connect such
quantities and thus they refer exclusively to the component form of tensor
quantities and in that form they transform only \emph{some} components of
the whole tensor quantity. In the ''AT relativity'' the quantities connected
by an AT are considered to be the same quantity, but such quantities are not
well defined in 4D spacetime, and actually they correspond to different
quantities in 4D spacetime.

The difference between the traditional ''AT relativity'' and the invariant
formulation of SR, i.e., the ''TT relativity,'' is also illustrated by the
difference in the interpretation of the Michelson-Morley experiment.

In Sec. 3 we have presented Maxwell equations as the true tensor equations (%
\ref{mxt3}) or (\ref{maxten}) and as the CBGEs (\ref{maco1}). It is
discussed how from these equations one finds the usual covariant Maxwell
equations (i.e., the component form) (\ref{maxco}).

In Sec.4 we have introduced the 4-vectors $E^{a}$ and $B^{a}$ instead of the
usual 3-vectors $\mathbf{E}$ and $\mathbf{B}$ and we have formulated the
Maxwell equations as tensor equations with $E^{a}$ and $B^{a},$ i.e., as the
CBGEs (\ref{maeb}) and the equations for the basis components $E^{\alpha }$
and $B^{\alpha }$ (\ref{ma4}) (all in the ''e'' coordinatization). These
equations are completely equivalent to the usual covariant Maxwell equations
in the $F^{ab}$- formulation, (\ref{maco1}) and (\ref{maxco}). It has been
explicitly shown in Sec. 4.1 that all the results obtained in a given IFR $S$
from the usual Maxwell equations with $\mathbf{E}$ and $\mathbf{B}$ remain
valid in the formulation with the 4-vectors $E^{a}$ and $B^{a}$ (in the
''e'' coordinatization), but only for the observers who measure the fields $%
E^{a}$\ and $B^{a}$\ and are at rest in the considered IFR. Thus we conclude
that the tensor quantities $E^{a}$ and $B^{a}$ do have the real physical
meaning and not the 3-vectors $\mathbf{E}$ and $\mathbf{B.}$ \bigskip

\noindent \textbf{ACKNOWLEDGEMENTS}\medskip

\noindent It is a pleasure to acknowledge to Professor Valeri Dvoeglazov for
inviting me to write this contribution for CONTEMPORARY\
ELECTRODYNAMICS.\bigskip


\end{document}